\documentclass[twoside]{article}
\usepackage{epsfig}

\input ijtaf.sty

\begin{document}

\runninghead{Michele Pasquini and Maurizio Serva}
{Indeterminacy in foreign exchange markets}

\normalsize\textlineskip
\thispagestyle{empty}
\setcounter{page}{1}

\fpage{1}
\centerline{\bf INDETERMINACY IN FOREIGN EXCHANGE MARKETS}
\vspace*{0.37truein}

\centerline{\footnotesize MICHELE PASQUINI \& MAURIZIO SERVA}
\vspace*{0.015truein}

\centerline{\footnotesize\it  Istituto Nazionale Fisica della Materia}
\baselineskip=10pt

\centerline{\footnotesize\it Dipartimento di Matematica, 
Universit\`a dell'Aquila }
\baselineskip=10pt
\centerline{\footnotesize\it I-67010 Coppito, L'Aquila, Italy}

\vspace*{0.225truein}
\pub{ \today }

\vspace*{0.21truein}
\abstracts{
We discuss price variations distributions in foreign exchange markets,
characterizing them both in calendar and business time frameworks.
The price dynamics is found to be the result of two
distinct processes, a multi-variance diffusion
and an error process. 
The presence of the latter, which dominates at short
time scales, leads to indeterminacy principle in finance.
Furthermore, dynamics does not allow for a 
scheme based on independent probability distributions, 
since volatility exhibits a strong correlation
even at the shortest time scales.\\
\vskip 0.2cm
JEL Category: C10
}{}{}

\vspace*{1pt}\textlineskip	

\section{Introduction and definition of time}	
\vspace*{-0.5pt}
\noindent

The characterization of price dynamics in financial markets
is an old but still puzzling problem.
At the beginning of the century, Bachelier~\cite{Bachelier} proposed
to consider price variations as independent realizations of identically
Gaussian distributed variables, while in the '60 Mandelbrot~\cite{Mandelbrot}
introduced symmetric L\'evy stable distributions. In the last years,
Mantegna and Stanley (\cite{MS1}, and see also \cite{BouchaudPotters,Stanley})
provided evidence that a L\'evy stable 
process  well reproduces the central part of high frequency price
variation distribution, while the tails are approximately exponential.
The Bachelier's Gaussian shape is recovered only 
on longer time scales which are, typically, of order one month.
The common point of
\cite{Bachelier}-\cite{Stanley}
is that price dynamics is considered as the result of 
independent random variables. This kind of approach seems inadequate,
since there is evidence of volatility correlations on long run 
\cite{Taylor}-\cite{PS2}.

Indeed, price distribution strongly depends on how one measures time flow. 
The choice of time index is twofold, calendar time and business time.
Business time is the sequence of integers $n=1,2,3,.....$ which 
indexes successive established quotes.
These integers correspond respectively to the calendar times 
$t_1,t_2,....$. Therefore, calendar time (which is monotonically increasing)
is a stochastic process of business time.
The relation can be inverted by considering $n$
as a function of calendar time, i.e. $n=n_t$,
but in this case the function is defined only on the sequence of 
quoting calendar times.
The price dynamics, therefore, can be described with respect 
to business time by means of the series $S_n$
or with respect to the calendar time by means of $S_t \equiv S_{n_t}$.
In the first cases quotes $S$ are defined on all integers and lags are 
all equal, in the second, they are defined only on quoting times 
and lags are unequal.

The main result of this paper is that a price quote is due to two distinct
independent stochastic processes: 
an {\it error} process superposed to an {\it underlying} price process.
The latter evolves following calendar time, while the former is 
due to an erroneous evaluation of a market operator, 
and its natural frequency is marked by business time. 
The resulting price variation distribution is, therefore, the convolution
of two distributions associated to these two distinct processes: 
the error process distribution, which does not change with  
time, and the distribution of the underlying process, which scales, on the
contrary, with calendar time.

The error process produces always a gap
between two consecutive price estimations even if they are almost
contemporary.
This phenomenology strongly reminds quantum mechanics, where a
measurement result always has a minimum uncertainty as stated by the 
Heisenberg Principle.
Following this comparison, one can state an {\it indeterminacy principle}
in finance: a price is never given with 
a precision less than a natural constant for that market.

Another important fact which is due to this phenomenology is that 
two consecutive price variations cannot be considered fully independent 
random variables, but they exhibit a very peculiar anticorrelation 
as we will see. 
Moreover, we provide evidence that volatility is so strongly correlated
that remains substantially constant inside the largest lag we consider,
and therefore the usual L\'evy stable scheme seems to be not appropriate.

In this paper we examine the high frequency 
price variation distribution of three foreign exchange markets, 
the Deutsch Mark/US Dollar (DEM-USD) exchange in 1993 (1,472,140 quotes)
and in 1998 (1,620,843 quotes), the Japanese Yen/US Dollar (JPY-USD)
in 1993 (570,713 quotes), and the Japanese Yen/Deutsch Mark (JPY-DEM)
in 1993 (158,878 quotes).
The quotes represent the value of one US Dollar in
Deutsch Marks and Japanese Yens in, respectively,
the DEM-USD and JPY-USD cases, while they represent the value of
one Deutsch Mark in Japanese Yens in the DEM-JPY case.
The price changes are given in pips, which indicate
a DEM/$10,000$ in the DEM-USD case, and a Yen/$100$ in the JPY-USD and
JPY-DEM cases.
All the data sets analyzed in this work 
have been provided by Olsen \& Associates.

We deal only with bid price, since we have found
more spurious data in the ask price distributions
due to wrong transcriptions. This is probably related to the fact that
in recording process quotes are specified by bid price and the 
last two digits of the bid/ask spread. 
Nevertheless, all the results can be fully reproduced if one deals with
ask price or average price.

The paper is organized as follows. 
In Section 2 it is shown that a price variation is given by the composition
of two distinct stochastic processes and that
the hypothesis of independent consecutive price
variations is incorrect, due to the presence of both price anticorrelation
and volatility correlation.
In Section 3 the price 
variation distribution for a single business lag and for the minimum 
calendar lag (two seconds) are computed. The latter turns out to be, 
basically, the error distribution.
In Section 4 we provide evidence that the underlying price variation
distribution at a generic calendar time lag is given
by a symmetric Gaussian process whose standard deviation
is itself a Log-normal process. 
In Section 5 some final remarks are reported.

\vspace*{1pt}\textlineskip	
\section{The indeterminacy principle in finance}
\vspace*{-0.5pt}
\noindent

The aim of this Section is to show that the set 
of reported bid quotes is a realization of stochastic process 
which is the composition of two processes 
whose origin and meaning is very different.
The first is the ordinary multiplicative stochastic process which
determines the evolution of underlying price,
while the second is a superposed stochastic noise which somehow accounts for 
the erroneous evaluation of the underlying price by 
the market operators. 
 
In order to demonstrate our point it is better
to consider the price evolution in business time. 

Let us assume that it exists an unobserved underlying 
price $\tilde{S}_n$ which accounts for
the real relative value of two currencies and which
follows the ordinary evolution rule

\begin{equation} 
\tilde{S}_{n+1} = \tilde{S}_n+ \tilde{R}_n
\label{true-price}
\end{equation}
where, as usual,  the $ \tilde{R}_n$ are, at the lowest approximation,
identically distributed variables which:   
  a) have vanishing average,
  b) only depend on the stochastic calendar 
lag $\Delta t_n \equiv t_{n+1}-t_n$,
  c) are uncorrelated, i.e. the average of the product
$\tilde{R}_n\tilde{R}_m$ vanishes for $n \neq m$,       
  d) their typical size (standard deviation) 
is proportional to $\tilde{S}_n$.
The last property it is necessary to ensure that the process is 
multiplicative on the large time scale.
Notice that we do not assume that the $ \tilde{R}_n$ are independent,
and the reason for that  will be clear at the end of this Section.
  
Let us also assume that the observed price 
(which is the recorded quote) slightly differs 
from the underlying price because of an erroneous 
evaluation of the operator.
The relation between the {\it underlying} and the {\it observed} 
price will be

\begin{equation} 
S_n = \tilde{S}_n + E_n
\label{obs-price}
\end{equation}
where the $E_n$ are zero mean, uncorrelated identically
distributed variables. Independence, again, is not assumed.
At variance with the $\tilde{R}_n$
they are independent from calendar lags $\Delta t_n$:
in fact they are a consequence of price evaluation process, 
and therefore they are present at each business time.

As a result of the joint action of this two processes
one has that after $m$ business lags the observed
price evolves according to

\begin{equation} 
S_{n+m}=S_{n} + E_{n+m} - E_{n} +
\sum_{i=0}^{m-1} \tilde{R}_{n+i}
\label{evolution}
\end{equation}

Now one can appreciate that price variation differently depends on
two contributions, the first, which represents the evolution of the underlying price,
is the sum of the $m$ variables, the second,
which represents the uncertainty inherent to the quoting operation,
is always the difference of {\bf only two} uncorrelated 
identically distributed variables.

If this theoretical framework is valid there should be 
various consequences. First, according to Eqn. (\ref{evolution})
 the variance of a price change after $m$ business time should be
\begin{equation} 
<(S_{n+m}-S_{n})^2>=2A+Bm
\label{variance}
\end{equation}
where $<\cdot>$ indicates the average over the probability distribution,
and $A= <E_n ^2>$ and $B=<\tilde{R}_n ^2>$.
 
This last average also runs on all possible calendar 
lags between two successive quotations.
The occurrence of this property can be appreciate in 
Fig. 1 which refers to DEM-USD 1998  exchange quotes,
and price variations are given in DEM/$10,000$. 
Results are highly consistent with the linear behaviour
with $A=5.9 \pm 0.2$ and $B=1.85 \pm 0.01$.

\begin{figure}
\vspace*{13pt}
\begin{center}
\mbox{\epsfig{file=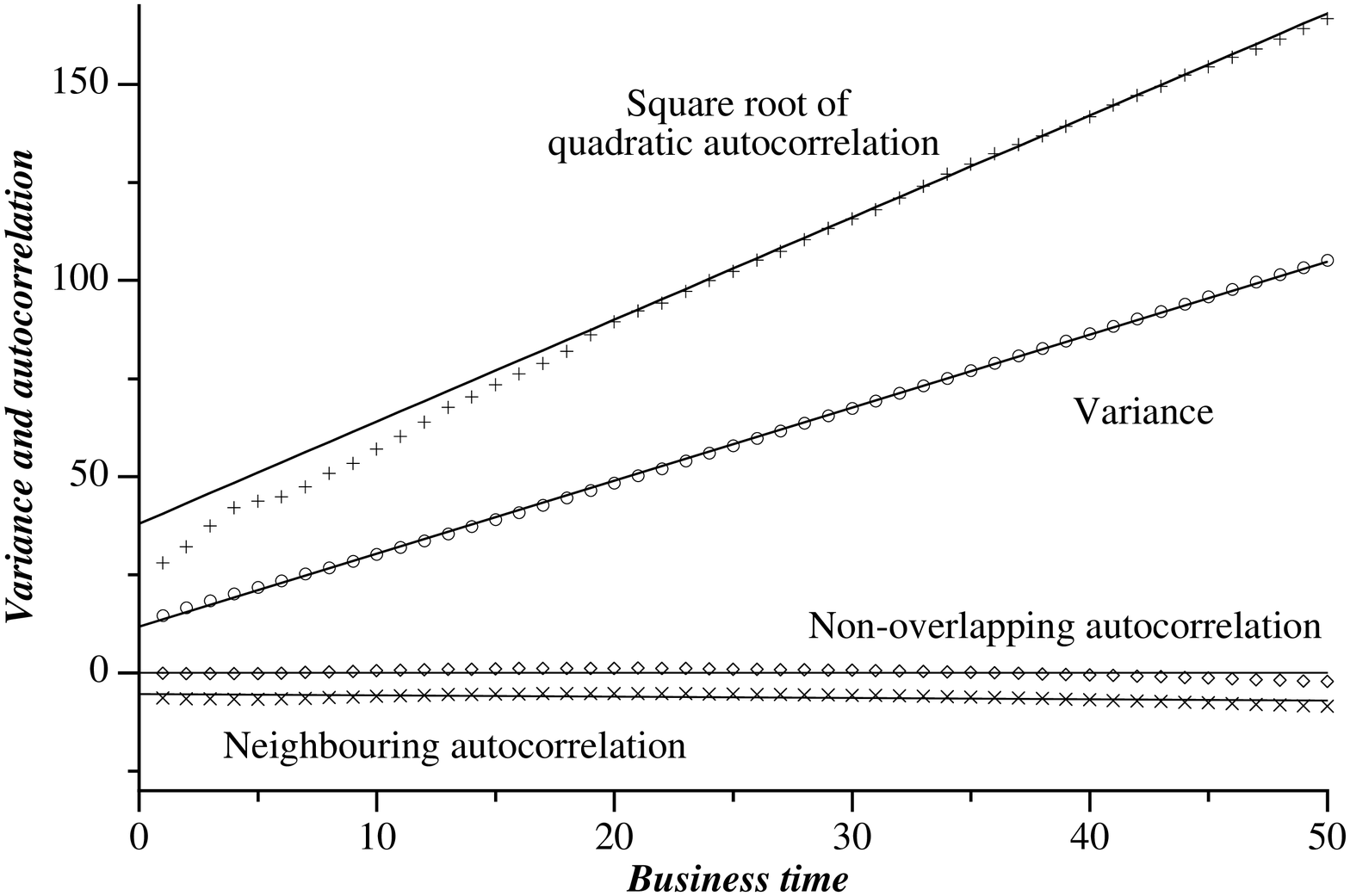,width=5.in}}
\end{center}
\vspace*{13pt}
\fcaption{ DEM-USD 1998 exchange rate: variance (\ref{variance}) (circles)
compared with a linear fit $2 A + B m$ with 
$A=5.9 \pm 0.2$ and $B= 1.85 \pm 0.01$,
neighbouring autocorrelation (\ref{corr-touch}) (slanting crosses)
compared with $-A$ from the previous linear fit,
non-overlapping autocorrelation (\ref{corr-untouch}) (rhombuses) 
compared with zero, and 
square root of quadratic autocorrelation (\ref{corr-vol}) (crosses)
compared with a scaling fit $C m$ in the range $20 \le m \le 50$
with $C=2.60 \pm 0.03 $.
}
\label{corr-dm98}
\end{figure}

The non-vanishing value of $A$ has an important meaning, since it
suggests that market price models based on continuous time approach
\cite{Merton} are not appropriate: in fact, this would at least require
in this case a vanishing price change in the limit of vanishing
lag. In this case, price change variance should be zero in the limit
$n \to 0$ (i.e. $A=0$), 
while in Fig. 1 it can be clearly appreciate the non occurrence
of this fact.

The second, more peculiar, consequence of Eqn. (\ref{evolution})
is that the neighbouring
autocorrelation of two consecutive price variations 
after $m$ business time is
\begin{equation} 
<(S_{n+m}-S_{n})(S_{n}-S_{n-m})>=-A
\label{corr-touch}
\end{equation}
where $A$ is the same parameter of (\ref{variance}).
The similar autocorrelation for non-overlapping 
price variations ($l > 0$), 
according to Eqn. (\ref{evolution}), are
\begin{equation} 
<(S_{n+l+m}-S_{n+l})(S_{n}-S_{n-m})>=0
\label{corr-untouch}
\end{equation}
Indeed, these last two equations represent a strong test in order 
to verify the goodness of the proposed price modelization, 
since phenomenology is very peculiar and crudely differs, for example, 
from Markov anticorrelations.

Both these relations are well satisfied as it can be appreciated in
Fig. 1. The same results can be found by using other data sets:
    $A=7.7 \pm 0.2$ and $B=2.59 \pm 0.01$ for DEM-USD 1993 exchange rate, 
    $A=8.4 \pm 0.4$ and $B=2.96 \pm 0.03$ for JPY-USD 1993, 
and $A=1.2 \pm 0.2$ and $B=4.88 \pm 0.01$ for JPY-DEM 1993.
In the last two cases price variations are given in Yen/$100$ (pips).

It should be noticed that in all cases $2 A$ is 
about ten times larger then $B$. This means that
the observed price changes are largely dominated by
the error effect on a short time scale.
Only after about ten business lags,
the underlying price variation is comparable with error
and it can be partially appreciated. Ten business lags correspond, 
approximatively, to two minutes in calendar time for the
DEM-USD cases. This means that price change distributions on
time scales of few minutes are deeply affected by
error, and do not reflect the real price variation.

At this point we are sufficiently convinced that
a price is a fuzzy variable, not only because of the obvious
bid/ask spread, but especially because the bid price itself
cannot be given at each time with an absolute precision less than $\sqrt{A}$.
We call this simple fact the {\it indeterminacy principle} of markets. 

In conclusion of this Section, let us consider that Eqn. (\ref{corr-untouch}),
which states that the autocorrelation of two non-overlapping
price variations is zero, may lead to the idea that these variables
are independent.
The inconsistence of this hypothesis can be provided by
computing the following non-overlapping quadratic autocorrelation:
\begin{equation} 
<(S_{n+m+1}-S_{n+1})^2(S_{n}-S_{n-m})^2> - <(S_{n}-S_{n-m})^2>^2
\label{corr-vol}
\end{equation}
If two non-overlapping absolute price variations were independent, 
one simply would have that this quantity vanishes.

On the contrary, the quadratic autocorrelation (\ref{corr-vol})
is sensibly different with respect to zero, as it can be appreciated in Fig. 1,
where its square root is plotted.
In fact, at large business times $m$, it shows a linear behaviour
$C \, m$. The numerical estimations give
$C=2.60 \pm 0.03$ for DEM-USD 1998 exchange rate,
$C=3.07 \pm 0.08$ for DEM-USD 1993,
$C=2.7  \pm 0.1 $ for JPY-USD 1993 and 
$C=4.2  \pm 0.3 $ for JPY-DEM 1993.

The last result gives an important information concerning 
the quadratic autocorrelation of underlying price process:
\begin{equation} 
<\tilde{R}_{n}^2 \tilde{R}_{n+m}^2> -
<\tilde{R}_{n}^2>^2 \simeq C^2
\label{corr-vol-un}
\end{equation}
independently on $m$ at least for business time lag $m$ 
in the range $20 \le m \le 50$
(about ten minutes for DEM-USD 1998 exchange rate). 
In other words, Eqn. (\ref{corr-vol-un}) asserts that absolute
price have an autocorrelation which is substantially independent
on the time separation $m$.
Indeed, this is a very surprising result, and
it confirms the presence of a strong correlation on 
absolute price changes not only on time scales from minutes \cite{DeJong}
to months \cite{PS1,PS2}, but also on shortest time scales of seconds.

Indeed, absolute price exhibit a very reach phenomenology also
for what concerns long term behaviour.
In particular, recent results on long memory \cite{PS1,PS2}
show power-law correlations up to time scale of one year for $|\Delta S|^x$, 
where $|\Delta S|$ is price variation and $x$ is a real number, 
whose exponent depends on $x$ (multiscaling).

\vspace*{1pt}\textlineskip	
\section{Single lag distribution and minimum calendar lag distribution}	
\vspace*{-0.5pt}
\noindent

According to (\ref{evolution}) the process 
changes are determined by the joint effect of underlying evolution and added noise.
The equation  (\ref{evolution}) can be easily rewritten in calendar time
as
\begin{equation} 
S_{t+\tau}=S_{t} + E_{n(t+\tau)} - E_{n(t)} +
\tilde{R}_{t,\tau}
\label{t-evolution}
\end{equation}
where $t$ and $t+\tau$ are two calendar times in which quotes are 
established, and $\tilde{R}_{t,\tau}$ represents the variation of
underlying price from time $t$ to time $t+\tau$ 
corresponding to the sum in (\ref{evolution}).

In consequence of (\ref{t-evolution}) 
and taking into account the uncorrelation of all the variables,
the probability distribution of a price change $R$ is given by the convolution

\begin{equation} 
P_\tau (R) = Q(\Delta E) \otimes \tilde{P}_\tau(\tilde{R})
\label{r-prob}
\end{equation}
where $Q(\Delta E)$ is the error distribution and $\tilde{P}_\tau(\tilde{R})$
is the probability distribution of underlying price $\tilde{R}$.
It should be noticed that
 $Q(\Delta E)$ is directly the probability of the differences
of the two variables $ E_{n(t+\tau)} - E_{n(t)}$ and 
not of a single one. 
The probability $Q(\Delta E)$ does not depend on the calendar
time lag $\tau$, at variance with $\tilde{P}_\tau(\tilde{R})$.

The error distribution $Q(\Delta E)$ and the underlying distribution
$\tilde{P}_\tau(\tilde{R})$ cannot be directly observed,
since the observable probability distribution is  only the $P_\tau (R)$.
Nevertheless, the analysis of the price change variance (\ref{variance})
suggests that at a single business time the distribution is largely
ruled by the error distribution.
Furthermore, since two seconds is the minimal calendar lag, 
($t_{n+1}-t_{n}$ is always a multiple of two seconds)
the corresponding price variations are a subset of those corresponding 
to the minimal business lag $n=1$.
At the light of these informations, we expected that 
the minimum calendar time lag distribution $P_\tau$
is practically equal to $Q$.

\begin{figure}[t]
\vspace*{13pt}
\begin{center} 
\mbox{\epsfig{file=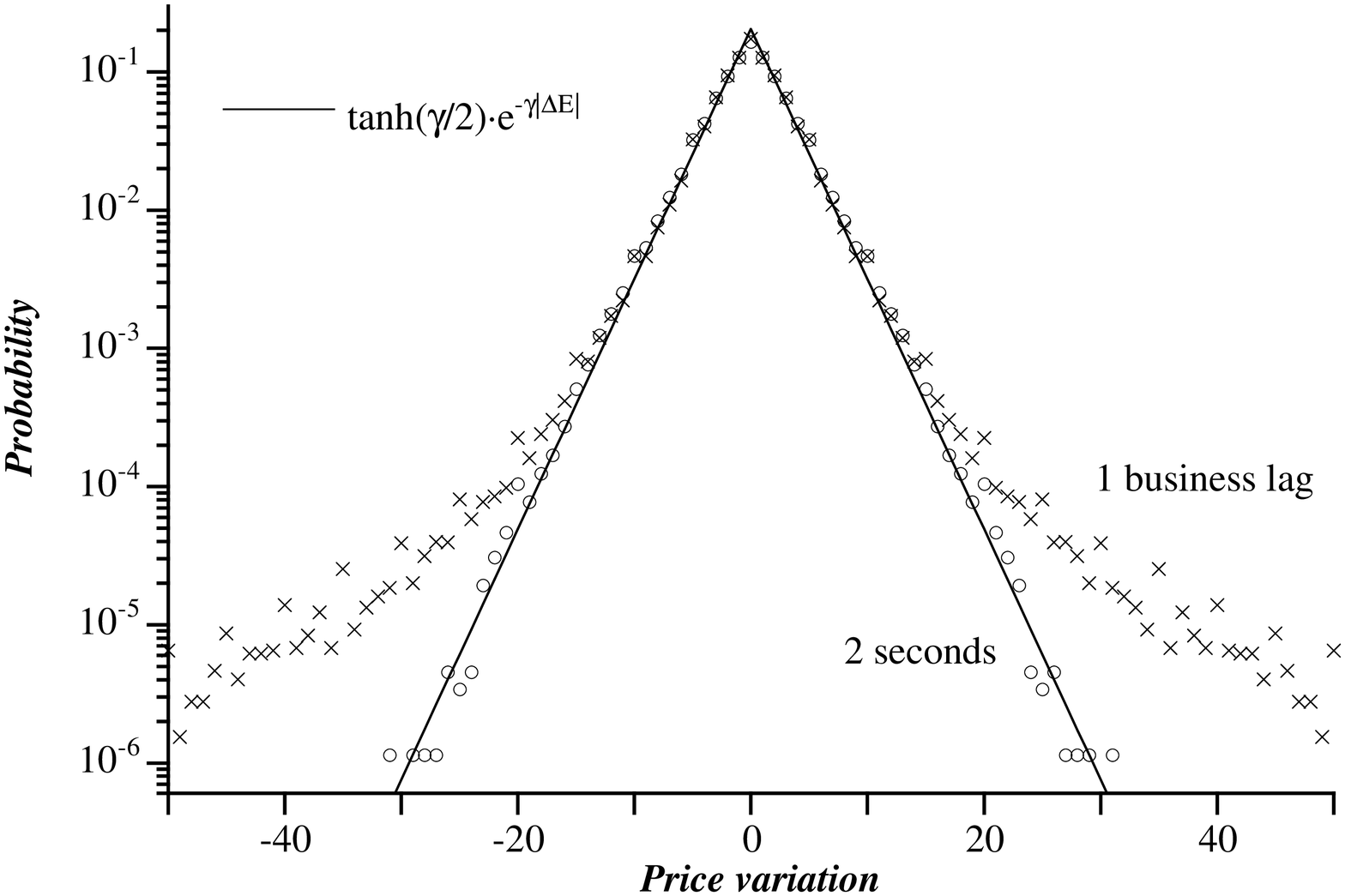,width=5.in}}
\end{center}
\vspace*{13pt}
\fcaption{
DEM-USD 1998 exchange rate: price change distribution at minimum calendar lag
of 2 seconds (circles) and at 1 business lag (crosses) in log-linear scales
as functions of price variation measured in pips (DEM/$10,000$).
The minimum calendar lag distribution basically coincides with the
error distribution $\tanh(\gamma/2) \exp(-\gamma|\Delta E|)$, 
and this gives the parametric fit $\gamma=0.40 \pm 0.01$ in the 
business time range $-23\le m\le +23$.}
\label{graf}
\end{figure}

In Fig. 2 we plot in log-linear scale
the two seconds calendar lag distribution.
The evidence of Fig. 2  is that the probability distribution $Q(\Delta E)$ is 
exponential of the form 
\begin{equation} 
Q(\Delta E) = \ \tanh{\gamma\over2} \ \exp(-\gamma|\Delta E|)
\label{e-prob}
\end{equation}
for {\bf six orders of magnitude}, and the numerical fit of the parameter
gives $\gamma=0.40 \pm 0.01$ in the business time range $-23\le m\le +23$.

A simple check of self-consistence can be made by computing the variance
of the error distribution $Q(\Delta E)$ and comparing the result
with $2 A$ of Eqn. (\ref{variance}). In fact, $Q(\Delta E)$ is
the probability distribution of the difference of two independent
variables $ E_{n(t+\tau)} - E_{n(t)}$, each one having variance
equal to $A$. After some algebra, the variance of distribution (\ref{e-prob})
reads:
\begin{equation}
< ({\Delta E})^2 > = {1\over \cosh\gamma - 1}
\end{equation}
By inserting into this formula the estimated $\gamma=0.40\pm0.02$, one has
$< ({\Delta E})^2 > = 12 \pm 1$, in agreement with the value 
we have found in the
previous Section $2 A = 11.8 \pm 0.4$.

In fig.2 the minimum business lag distribution is also plotted,
for comparison with the two seconds distribution.
The central parts of the two distributions are substantially the same
for three orders of magnitude, in this region the error process rules. 
Outside the region the single business lag distribution
exhibits more persistent tails.  These large events, 
corresponding to calendar time lags larger than two seconds, 
are due to underlying price variations instead of erroneous price changes. 

\vspace*{1pt}\textlineskip	
\section{Distributions at different calendar time lags}	
\vspace*{-0.5pt}
\noindent

In this Section we discuss the probability distributions of
price variations
for calendar lags of length $\tau$ from its minimum of two seconds
up to time scales of order of few minutes. 
In particular, we try
to retrieve informations about the underlying price distribution
$\tilde{P}_\tau(\tilde{R})$ in Eqn. (\ref{r-prob}). 

A possible strategy is to guess a functional parametric
expression for $\tilde{P}_\tau(\tilde{R})$, and then to compare its
$Q(\Delta E)$-convolution (\ref{r-prob}) 
with our experimental probability distributions.
In order to make a reasonable hypothesis on $\tilde{P}_\tau(\tilde{R})$, 
let us come back to business time approach
and write down the underlying price change $\tilde{R_n}$ 
as the product of a volatility $\sigma_n$ and a Normal Gaussian variable 
with zero mean and unitary variance
$W_n$:
\begin{equation}
\tilde{R_n} \ = \ \sigma_n \ W_n
\end{equation}
Notice that this decomposition always allows for independent
Normal variables $W_n$.
From Eqn. (\ref{corr-vol-un}) and independence of $W_n$,
one can derive a similar relation
for the volatility:
\begin{equation}
<\sigma_n^2  \sigma_{n+m}^2> - <\sigma_n^2>^2 \simeq C^2
\end{equation}
for $m$ up to 50.

The simplest interpretation of the last result is that volatility
remains substantially constant in business time lags of length $m$, 
but it assumes different values for temporally far lags 
(otherwise the last expression should vanish).
Than $C^2 \simeq <\sigma_n^4>-<\sigma_n^2>^2$.
In other words, in a business lag of $m$ steps the volatility process is
substantially frozen, while one has $m$ independent realizations of
the Gaussian process.
This result identically holds for  calendar time,
where one says that in a time lag of length $\tau$ up to, about, 
ten minutes for DEM-USD 1998 exchange rate,
one has a constant volatility. This volatility can 
obviously change with respect
to that of another time lag of length $\tau$ temporally far.

In other terms, volatility itself is a stochastic process
with a characteristic time much larger than about ten minutes.
The underlying price variation $\tilde{R_{t,\tau}}$ on a calendar lag of length $\tau$
is then the sum of several independent Gaussian variables with same variance.
Therefore, it can be written as:
\begin{equation}
\tilde{R_{t,\tau}} \ = \ \sigma_t \sqrt{\tau} W_t
\end{equation}
where $\sigma_t$ is constant in the calendar lag $(t,t+\tau)$
and $W_t$ is a Normal Gaussian random variable.

The price process framework is completed when
an explicit expression for the volatility probability distribution
is given.
With this aim, let us recall recent results \cite{Stanley,PS2}, 
where it was found that returns
probability distributions of a stock market daily index 
and of a foreign daily exchange are symmetric Gaussian distributions 
whose standard deviations are themselves Log-normal processes.
Borrowing this result and taking into account the previous considerations,
the following expression for $\tilde{P}_\tau(\tilde{R})$ comes out:

\begin{equation}
\tilde{P}_\tau(\tilde{R}) = 
\int_0^{\infty}  d\sigma \ L(\sigma) 
\int_{\tilde{R}-{1\over2}}^{\tilde{R}+{1\over2}}  dr 
\ G_{\sigma\sqrt{\tau}}(r)
\label{p-guess}
\end{equation}
where $L(\sigma)$ is a Log-normal distributions of parameters 
$\mu$ and $\omega$
\begin{equation}
L(\sigma) = \ {1\over{\omega \ \sqrt{2 \pi}}}
\ \exp \left[ -{(\ln \sigma - \mu)^2}\over{2\ \omega^2} \right]
\label{lognorm}
\end{equation}
and $G_{\sigma\sqrt{\tau}}(r)$ is a Gaussian distribution with zero mean
and standard deviation $\sqrt{\tau}\sigma$, integrated on the unitary
interval centered on the integer $\tilde{R}$.

We have performed a numerical computation of Eqn. (\ref{p-guess}) 
via a MonteCarlo approach, and then we have convoluted the result
with the error distribution $Q(\Delta E)$. 
For the DEM-USD 1998 exchange rate we have found a parameters choice
($\mu=-1.5$ and $\omega=0.8$) which gives a good agreement 
between the guessed and the experimental price change distributions,
as shown in Fig. 3, for three values of time lag $\tau$: $4$, $40\pm 2$ 
and $400\pm 20$ seconds. In order to have a reasonable number of occurrences,
the experimental distributions are computed 
with a $5\%$ tolerance on $\tau$.

\begin{figure}[t]
\vspace*{13pt}
\begin{center} 
\mbox{\epsfig{file=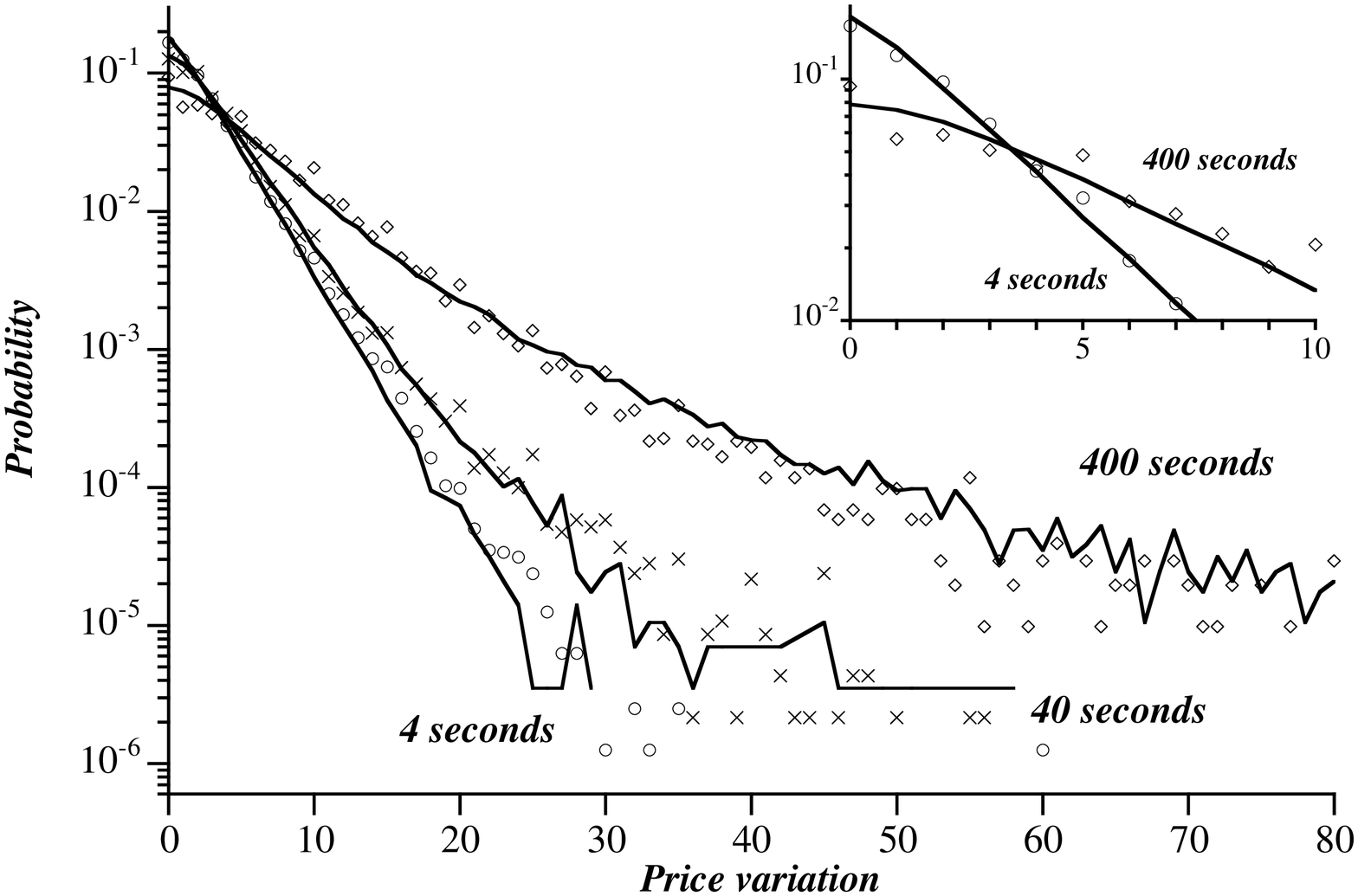,width=5.in}}
\end{center}
\vspace*{13pt}
\fcaption{
DEM-USD 1998 exchange rate: price change distributions for calendar time lag
$\tau$ equal to $4$ seconds (circles), $40\pm2$ seconds (crosses) 
and $400\pm20$ seconds (rhombuses) in log-linear scales
as functions of price variation measured in pips (DEM/$10,000$).
The lines represent the numerical computation of the guessed distribution
$\tilde{P}_\tau(\tilde{R})$
(\ref{p-guess}) convoluted with $Q(\Delta E)$ (\ref{e-prob}) for $\mu=-1.5$ and
$\omega=0.8$.
In the inset the small price changes range is magnified for the cases
4 seconds and 400 seconds.}
\label{distr}
\end{figure}

The consistency of the hypothesis of a Log-normal volatility which is 
constant during a whole time lag
can be directly checked computing 
$\sqrt{<\sigma_n^4>-<\sigma_n^2>^2}$ for such a distribution and comparing
the result with the experimental value of $C$. In the DEM-USD 1998 case
one has that the first computation gives 
$\simeq 2.77$ against the value $C=2.60 \pm 0.03$.
Therefore the constancy of volatility
seems reasonably confirmed.

The conclusion is that the underlying volatility itself evolves 
following a stochastic process, with characteristic times
larger than about ten minutes, while on shorter time scales the
snapshot variance is mainly due to the error process. 
The hypothesis of a Log-normal underlying volatility which
seems once more confirmed by experimental data,
 could be the result of a multiplicative process.

\vspace*{1pt}\textlineskip	
\section{Conclusions}
\vspace*{-0.5pt}
\noindent

We have found evidence that two distinct stochastic processes, an
underlying process
and an error process, are present in price change dynamics in foreign 
exchange markets. The latter process, which can be put in connection with
the price estimations of market operators, suggests a quantum like
nature for price
changes variables, in the sense that they have 
an irreducible intrinsic indeterminacy.
For this reason market models based on continuous time limit seem to be not
adequate.

Another main consequence of the indeterminacy principle is that price changes
cannot be considered the result of independent stochastic processes,
because of the presence of correlations even at the shortest time scales. 
In fact, the spurious anticorrelations 
in the observed price changes \cite{BPSVV}
can be easily explained in the light  of the error process.

At this point it seem quite clear that a moving chart over few
observed prices is able to give a more accurate estimations of the
underlying price at a given time. Nevertheless,
the number of observed prices in the moving chart cannot be too large
(i.e. calendar time lag cannot be too long) otherwise
too large price variations due to the underlying process
would decrease accuracy.

We expect to find effects of the indeterminacy
principle also in high frequency market quotes of single
stocks. On the contrary, the error component in change distributions
of composite market indexes are probably not so important 
because of the averaging of errors due to 
the large number of different stocks involved.

Finally, the volatility of the underlying process exhibits 
a strong autocorrelation, since it turns out to be substantially
constant up to time scales of, at least, ten minutes. 
This is a clear evidence
that the underlying volatility is a stochastic process
with a larger characteristic time, while variance 
at short time scales is basically due to the error
process. We have proposed that the underlying volatility follows
of a Log-normal process and that the underlying price change distribution
is a symmetric Gaussian whose standard deviation
is the Log-normal stochastic volatility. 
This hypothesis is in a very good agreement with experimental data.

\bigskip
\noindent
{\bf Acknowledgment}

\noindent
We thank Roberto Baviera
whose comments and suggestions have greatly contributed 
to realization of this work.

\end{document}